\documentclass{article}
\usepackage[margin=1in]{geometry}
\usepackage[square,sort&compress,comma,numbers]{natbib}

\usepackage{graphicx}
\usepackage{float}

\title{Wind dispersal of natural and biomimetic maple samaras}

\author{Gary K. Nave, Jr.$^1$\thanks{Corresponding author: {\tt gary.k.nave@gmail.com}} \and Nathaniel Hall$^1$ \and Katrina Somers$^2$ \and Brock Davis$^3$ \and Hope Gruszewski$^2$ \and Craig Powers$^4$ \and Michael Collver$^5$ \and David G. Schmale III$^2$ \and Shane D. Ross$^6$}

\date{
    \raggedright
1 Engineering Mechanics program, Virginia Tech, Blacksburg, Virginia, USA \\
2 School of Plant and Environmental Sciences, Virginia Tech, Blacksburg, Virginia, USA \\
3 Department of Mechanical Engineering, Virginia Tech, Blacksburg, Virginia, USA \\
4 Department of Civil and Environmental Engineering, Virginia Tech, Blacksburg, Virginia, USA \\
5 Blacksburg High School, Blacksburg, Virginia, USA \\
6 Department of Aerospace and Ocean Engineering, Virginia Tech, Blacksburg, Virginia, USA \\
}

\begin{document}

\maketitle

\section*{Abstract}
Maple trees (genus \emph{Acer}) accomplish the task of distributing objects to a wide area by producing seeds which are carried by the wind as they slowly descend to the ground, known as samaras.
With the goal of supporting engineering applications, such as gathering environmental data over a broad area, we developed 3D-printed artificial samaras.
Here, we compare the behavior of both natural and artificial samaras in both still-air laboratory experiments and wind dispersal experiments in the field.
We show that the artificial samaras are able to replicate (within 1 standard deviation) the behavior of natural samaras in a lab setting.
We further introduce the notion of windage to compare dispersal behavior, and show that the natural samara has the highest mean windage, corresponding to the longest flights during both high wind and low wind experimental trials.
This research provides a bioinspired design for the dispersed deployment of sensors and provides a better understanding of wind-dispersal of both natural and artificial samaras.

\noindent{\it Keywords\/}: wind dispersal, maple samaras, autorotation, additive manufacturing, biomimicry

\section{Introduction}
Distributed networks of inexpensive sensors can provide an effective method for gathering environmental data, with applications to precision meteorology and atmospheric physics \cite{chilson2019moving,barbieri2019small,lee2019use}, wildfire management \cite{rabinovich2018toward,al2019review}, air quality \cite{villa2016overview,carranza2018vista}, water quality \cite{schmale2019perspectives}, agricultural management \cite{hart2006environmental,barrientos2011aerial}, and even exploration of other planetary surfaces \cite{pandolfi2013biomimetics}. 
While much research focuses on sensor development and distributed sensor networks \cite{lohr2010smart,batista2017}, there remains a need for efficient methods to distribute the sensors themselves. 
For instance, in recent years, a concept known as GlobalSense has been in development which would involve massively deployable, low-cost airborne sensors inspired by two-winged seeds for atmospheric characterization \cite{prather2016,horton2018,bolt2020}.

Wind dispersal is a common distribution strategy in the biological world, employed by organisms across different scales, from microscopic scales (e.g., fungi and pollen) to macroscopic scales (e.g., insects and plant seeds \cite{howe1982ecology,hughes1994predicting,Isard2001,SchmaleRoss2015}).
%Prior research has shown that certain shapes naturally exhibit complex motions as they descend through the atmosphere. By choosing to exploit these complex behaviors of descent, we can control the landing distribution of a group of sensors with minimal energy input. The proposed work will use analytical tools and experimental validation to develop such a technology.
The fruit of maple trees (genus {\it Acer}) produce wind-dispersed seeds known as samaras. % (commonly referred to as ``helicopter seeds).
A samara consists of a seed (nut or pericarp) and a single fibrous wing.
After abscission, the  samara, after falling from rest, goes through a transition phase (about 1 m descent) which leads to  autorotation and a steady descent, slower than what would be predicted from the terminal velocity based on mass and drag considerations alone. 

Detailed fluid experiments have revealed the slow descent of the maple samara is due to a lift force generated by the formation of a stable leading-edge vortex \cite{lentink2009leading}, which has been reproduced in simulation via computational fluid dynamics \cite{lee2018scaling}.
%Biologically, the slow descent time provides a survivability advantage .
Samaras with a longer descent time have a greater likelihood of being dispersed over greater distances by the wind. This ultimately reduces resource competition and increases fitness \cite{fenner2012seed}.

In this study, we sought to reproduce the flight behavior seen in natural samaras by creating artificial samaras. 
We hypothesized that the artificial samaras would have similar dispersal characteristics (e.g., descent speed, rotation speed, and dispersal distance) to natural samaras.
To test this hypothesis, we developed replicas of samaras via additive manufacturing (i.e., 3D printing). 
These artificial samaras were designed to match the dimensions, shape, and weight distribution %center-of-mass characteristics 
of natural maple samaras. 
%We hypothesized that approximately matching the shape would lead to similar dynamical characteristics in both a laboratory (quiescent) setting and a field (windy) experimental setting, e.g., descent speed, rotation speed, and dispersal distance. 
We then conducted a series of laboratory (no wind) and field (with wind) experiments to study the dispersal of artificial and natural samaras. 
To understand the dispersal of the samaras by wind, we introduce
%The field experiments allow the consideration of how samaras travel along the wind.
%To quantify this effect, we introduce 
the concept of windage, representing the  horizontal force of the wind on, and hence the horizontal motion of, falling objects. %due to the direct exposure to wind forcing.
Windage is known to influence the fate of seeds of seagrasses, which are primarily water-dispersed \cite{harwell2002long,ruiz2012role,ruiz2015contemporary}, but to our knowledge has not been considered for primarily wind-dispersed seeds.

Artificial samara seeds, not of maples, but of a tropical tree
{\it Tachigalia versicolor}, have been considered previously \cite{augspurger1987wind}.
These artificial samaras autorotated, but about two axes simultaneously; the vertical axis and a longitudinal axis, much like the poplar seed (of the {\it Salicaceae} family) \cite{minami2003various}. 
By contrast, our artificial seeds only autorotate about a single vertical axis, like natural seeds of the maple genus ({\it Acer}). 
In the previous artificial samara study \cite{augspurger1987wind}, seeds were drop tested in the field, and their landing distributions recorded, but the wind speeds were not measured and the windage was not determined.

To reiterate, the specific objectives of our study were to: (1) create artificial samaras with a {\it morphology} similar to natural samaras and (2) compare the resulting {\it dynamical properties} of artificial samaras with natural samaras by conducting identical experiments on both sets of objects.

\hspace{5mm}

\section{Development of artificial samaras}

\begin{figure}
        \centering
        \includegraphics[width=\textwidth]{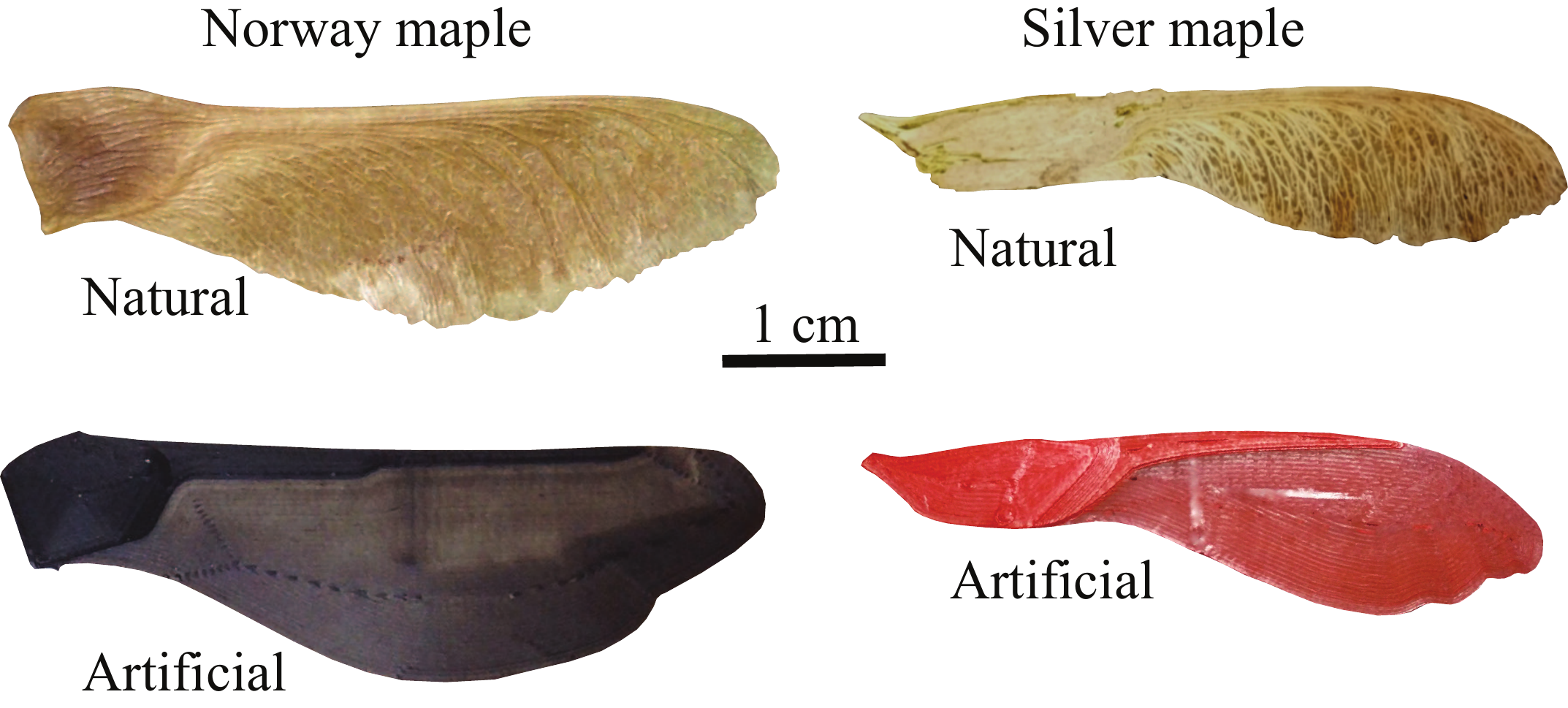}
        \caption{For both {\it Acer platanoides} (Norway maple) and {\it Acer saccharinum} (silver maple), we show the natural samara (seed) and an artificial, 3D-printed version.}
        \label{fig:development}
\end{figure}

We considered two species,
{\it Acer platanoides} (Norway maple) 
and
{\it Acer saccharinum} (Silver maple). 
We based the shape of a 3D-printed replica on the planform of one individual ($n=1$) of each species.
%The primary novel contribution of the present work is the design and testing of simple, 3D-printed artificial samaras, printed using an Ultimaker dual extrusion printer.
% To develop the model, we used reference images, such as the one in Figure \ref{fig:development}(a) of existing samaras to reconstruct the geometry.
Using reference points and spline fitting for a Norway maple and silver maple, we were able to reproduce the geometry of samaras in three 2-dimensional sections --- the nut, the leading edge, and the wing.
We extruded each samara section in the perpendicular direction and rounded all edges to generate the 3D-model for printing. 
The natural samara and its artificial 3D-printed counterpart is shown for each samara type in
figure \ref{fig:development}. 
%shows the use of the reference image and spline modeling, while further design details may be seen in the supplementary 3D-models.

We used an Ultimaker 3 3D printer with PLA (polylactic acid) as filament, which has a density of 1.24 g/cm$^3$.
Our 3D-printed samaras went through several design iterations while attempting to recreate the autorotation behavior of the natural samaras, varying both nut thickness and wing thickness. 
Due to the technical challenges of 3D-printing a thin layer of material, a wing thickness of 0.05 mm was %ultimately 
chosen, with a nut thickness of 1.5 mm heuristically providing the desired autorotation properties.

Due to seed availability for testing, samaras of the Norway maple ({\it Acer platanoides}) were chosen as 30 individuals ($n=30$) were available for further testing.
In all, 30 
%alongside the initial silver maple seeds, and drop tests were conducted on real 
natural Norway maple seeds, 30 artificial 3D-printed Norway maple seeds, and 30 artificial 3D-printed silver maple seeds were used in the experiments described below.
% These nut and wing thickness values were used for both 3D-printed Norway maple and silver maple seeds.

\section{Experimental methods}

\subsection{Samara morphology}
% Nature has optimized the Norway Maple over time to achieve the best performance for its goals.
There are several aspects of the maple seed that give it its autorotation abilities, such as a thick leading edge compared to the wing, and a dense seed that optimizes the center of mass for autorotation. 
To capture this morphology of the maple seed, we used a sectioning method to obtain a linear mass density along the long axis of the seed.
% a Micro-CT scanner (need to acquire model/type) was used to image the seed.
% The X-ray images obtained displayed the differing density throughout the seed and provided an geometrically accurate 3D model of a maple seed, as seen in Figure.... 
% 32 maple seeds were imaged and stored and available at … ? 
% Although a geometrically accurate 3D model for biomimicry was obtained, the model did not represent the changing density throughout the seed,and the proper software to extract this from the CT images was not available at the time.
% To further the later analysis of flight performance for the maple seeds, another method was used to get density distribution throughout the samaras. 
Although each maple seed is unique in its shape, mass, and wing style, a common  characteristic obtained from the linear density is the spanwise position of the center of mass. 
To approximate the linear density and thus the center of mass location, each of the 30 Norway maples used in dynamical testing were sliced into segments ranging from 2 mm to 7 mm (see figure \ref{fig:density}(a)), and each segment weighed using a 1 microgram precision scale. 
Using this mass and length data, the center of mass was calculated for each samara. %; figure \ref{fig:density}(b).
As this measurement requires destruction of the samara, this analysis was done after both the laboratory (still air) and field experiments.

\subsection{Still air experiments}
To determine the natural descent properties of all samaras tested, artificial and natural, we dropped the samaras in a still air setting, as in \cite{varshney2012kinematics}.
Samaras were released from rest, with the blade pointing downward, from a height of 3.05 m, and allowed to fall freely.
Tests were conducted in a laboratory (a growth chamber) without active ventilation, to minimize air currents that could disrupt the measurements.
Each individual samara 
%30 Norway maple seeds, 30 3D-printed Norway maple seeds, and 30 3D-printed silver maple seeds, 
was tested 3 times (i.e., 90 samaras, each dropped 3 times, giving 270 total drops).
A Photron FASTCAM Mini UX100 camera (Photron, San Diego, CA, USA) with a micro-Nikkor 105 mm f/28 lens (Nikon, New York, NY, USA) was used to record video data at 2000 frames per second (0.5 ms between frames) with a resolution of 1,280 $\times$ 800 pixels. 
The resulting images were analyzed to determine the mean descent speed and rotational velocity at steady state, as illustrated in figure \ref{fig:stillAirData}(a). 
Frames were processed using \texttt{opencv} for Python to identify samara position and orientation.

\subsection{Field experiments}
For a more realistic test of the samaras' descent performance, experiments were conducted in field conditions.
Samaras were dropped from an aerial work platform over an asphalt airstrip located at the Kentland Experimental Aerial Systems 
%(KEAS) 
Laboratory  at Virginia Tech's Kentland Farm in Blacksburg, VA (figure \ref{fig:ODscatter}(c)).
This site was selected because the paved area and the land in the immediate vicinity are flat, with no obstructions.
As in the still air experiments, samaras were released from rest from a height of $h=$ 3.05 m with the blade pointing downward and allowed to freely descend to the ground. The location of their first contact with the ground was marked with a colored metal disk so that no additional lateral sliding movement along the ground would occur.

Each natural Norway maple seed was dropped 2 times and each 3D-printed Norway maple seed and 3D-printed silver maple seed was dropped 3 times from the platform.
In this experiment, the repeated trials were conducted on different days to account for potential variability in wind conditions.
Wind conditions were approximated by taking an average for the entire trial duration, which was typically 3 hours per trial, on three different days. 
Wind speed and direction were measured at an approximate height of 3 m at a weather station located approximately 100 m from the drop location.
The wind conditions were variable. 
On the first day, artificial silver maple seeds were dropped in low wind conditions ($\bar{u}_w=0.95$ m/s) and artificial Norway maple seeds were dropped in higher wind conditions ($\bar{u}_w=3.69$ m/s).
On the second and third days, all samaras were dropped together, with one low ($\bar{u}_w=0.96$ m/s) and one higher ($\bar{u}_w=3.47$ m/s) wind day.
The wind conditions for each drop period are shown in Figure \ref{fig:ODscatter}(b).

\subsection{Windage}

%From the field drop data, the windage can be experimentally estimated. 
Windage was estimated from the field experiments.
We use the widely used and simplistic (kinematic) empirical model of windage \cite{ruiz2012role,allshouse2017impact,breivik2011wind}, which holds that the average horizontal velocity of the object, $u$, is linearly related to the average horizontal wind velocity $u_w$,
\begin{equation}
        u = C_w u_w,
\end{equation}
by a windage coefficient $C_w$, where $u_w$ is the horizontal wind velocity averaged between the release height, $h$, and the ground. To estimate $C_w$, we determine the downwind distance traveled by the dropped samara, $d$, as compared with the distance, $d_w$, traveled by a theoretical tracer following the horizontal wind for the same drop time.
Using each day's average wind speed, $\bar{u}_w$ and the drop time of each samara, $t$, the effective distance traveled by a passive tracer in the wind may be calculated as,
\begin{equation}
        d_w = \bar{u}_w t.
\end{equation}
The windage coefficient, $C_w$, therefore, is the ratio of the actual downwind distance traveled to the effective distance of a passive tracer,
\begin{equation}\label{eq:cw}
        C_w = \frac{d}{d_w}.
\end{equation}
Ignoring turbulence \cite{okubo1989theoretical,murren1998seed}, the drop time is approximately related to the release height and terminal velocity of the seed, that is, the seed descent rate in still air ($v_d$),
\begin{equation}
        t = \frac{h}{v_d},
\end{equation}
thus, the distance traveled is estimated as,
\begin{equation} \label{distance_windage}
        d =C_w h \frac{u_w}{v_d}.
\end{equation}
We note that this differs from a simple ballistic model used in some previous seed dispersal studies \cite{cremer1977distance,greene1989model,murren1998seed}, which effectively considered a windage coefficient of 1.

\section{Experimental results}

The mean and standard deviation of all measured properties are  shown in Table \ref{tab:properties}.

\begin{table*}[h]
\begin{tabular*}{\textwidth}{@{\extracolsep{\fill}} | l || c | c | c |}
\hline
Samara &  Natural Norway maple & Artificial Norway maple & Artificial silver maple \\ \hline \hline
 Mass, $m$ & $102.4 \pm 30.8$ mg & $183.6 \pm 2.3$ mg & --- \\ 
 Length, $L$ & $5.65 \pm 0.76$ cm & $5.00 \pm 0.00$ cm & --- \\ 
 Center of mass & $28.5\pm4.0$\% & $27.7\pm1.2$\% & --- \\ 
 Wing loading, $W/S$ & $2.09\pm0.50$ N/m$^2$ & $3.18\pm0.13$ N/m$^2$ & --- \\
 Rotational velocity, $\Omega$ &  $81.4\pm27.6$ rad/s & $93.7\pm14.4$ rad/s & $138.6\pm17.7$ rad/s \\
 Wing tip speed, $v_t=R \Omega$ &  $3.20\pm0.94$ m/s & $3.36\pm0.33$ m/s & --- \\
 Lab descent speed, $v_d$ & $1.10 \pm 0.24$ m/s & $1.28 \pm 0.17$ m/s & $1.36 \pm 0.22$ m/s \\
 Field descent speed, $v_d$ & $1.03 \pm 0.32$ m/s & $1.29 \pm 0.36$ m/s & $1.32 \pm 0.36$ m/s \\
 Windage coefficient, $C_w$ & $0.87\pm0.39$ & $0.78\pm0.33$ & $0.81\pm0.38$\\
 \hline
 \end{tabular*}
\caption{Morphological and dynamical properties of natural Norway maple, 3D-printed Norway maple, and 3D-printed silver maple samaras across all three sets of experiments: morphological, laboratory, and field testing. The mean and standard deviation are reported for each characteristic. For the 3D-printed silver maple, the mass, center of mass, and wing loading were not measured.}
\label{tab:properties}
\end{table*}

\subsection{Samara morphology}\label{ss:morphology}
\begin{figure}
\begin{center}    
	\begin{tabular}{c@{\hspace{0.5pc}}c}
    \includegraphics[height=0.33\linewidth]{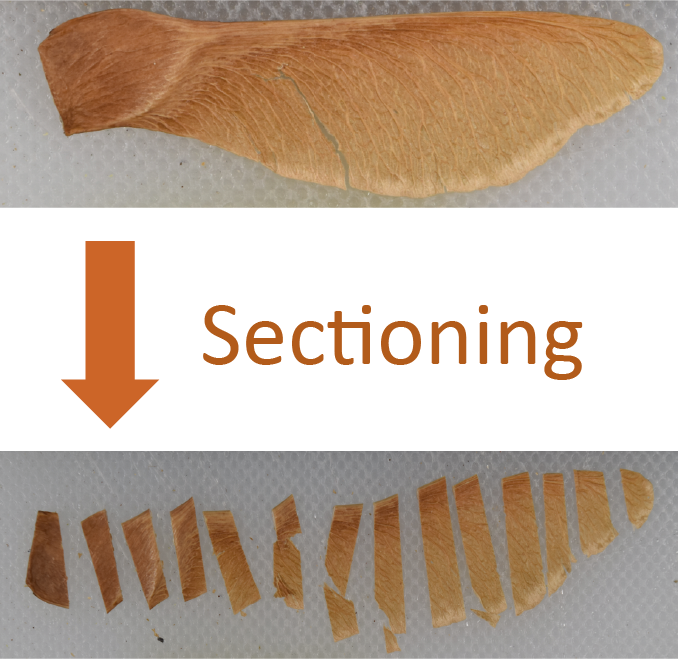}
     &
    \includegraphics[height=0.33\linewidth]{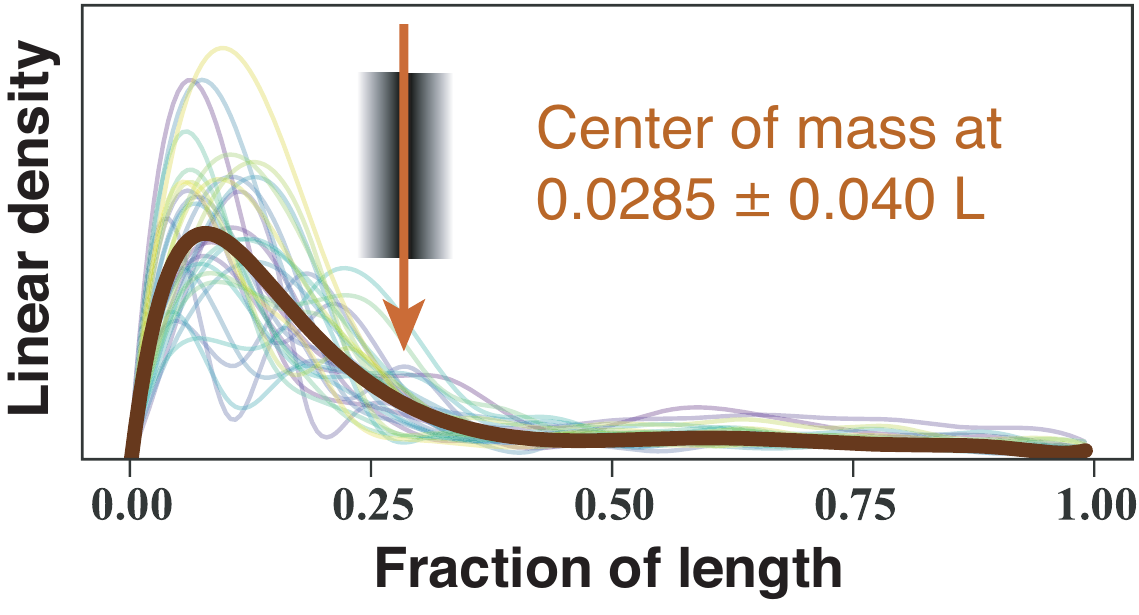} \\
    (a) & (b)
    \end{tabular}
	\caption{
	(a) A Norway maple seed ({\it Acer platanoides})  is sliced into sections and weighed in order to estimate the spanwise linear mass density (b), shown on a relative scale for 30 natural samaras, as well as an average (thick line).  With the spanwise length denoted as $L$, the center of mass was found to be located slightly more than a quarter of the way from the heavy (seed) end, more precisely, at 28.5 $\pm$ 4.0\hspace{0mm}\% $L$.
	}
	\label{fig:density}
\end{center}	
\end{figure}

The natural Norway maple samaras have a center of mass that occurs between 24\% and 30\% of its length measured from the heavy tip of the seed.
The average center of mass was 28.5 $\pm$ 4.0\%, as illustrated for a seed of {\it Acer platanoides} (Norway maple) in figure \ref{fig:density}(b).

The wing loading was also calculated for each of the 30 samaras,  with the average wing loading found to be 2.1 N/m$^2$. % , which agrees with Lentink's figure and analysis
Similarly, the artificial  Norway maples were sectioned and weighed to analyze their mass distribution characteristics.
Their center of mass was found to be 27.7\%, with a wing loading of 3.2 N/m$^2$.
As their natural counterparts were not available for testing in large enough numbers, the artificial silver maples did not undergo morphology characterization.
The artificial Norway maple seeds were found to be substantially heavier ($183.6\pm2.3$ mg) than their natural counterparts ($102.4 \pm 30.8$ mg), leading to a larger wing loading (artificial: $3.18\pm 0.13$ N/m$^2$, natural: $2.09\pm 0.50$ N/m$^2$).
However, the center of mass of artificial Norway maple seeds ($27.7\pm1.2$\%) was found to be very consistent with the natural seeds ($28.5\pm4.0$\%).

\subsection{Still air experiments}
\begin{figure}
    \centering
	\begin{tabular}{ccc}
    \includegraphics[height=0.27\linewidth]{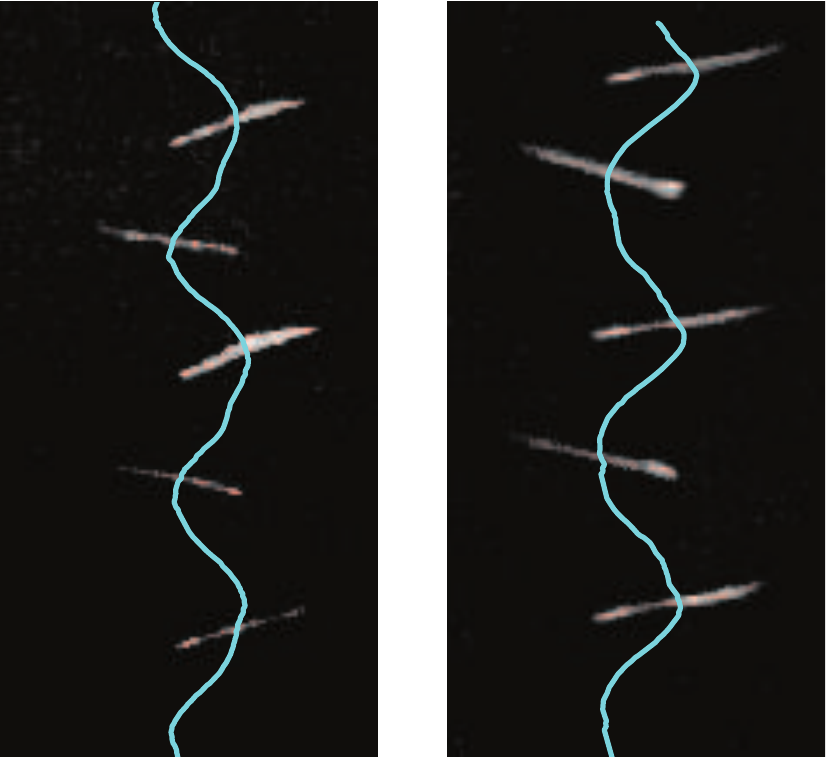}
    &
    \includegraphics[height=0.27\linewidth]{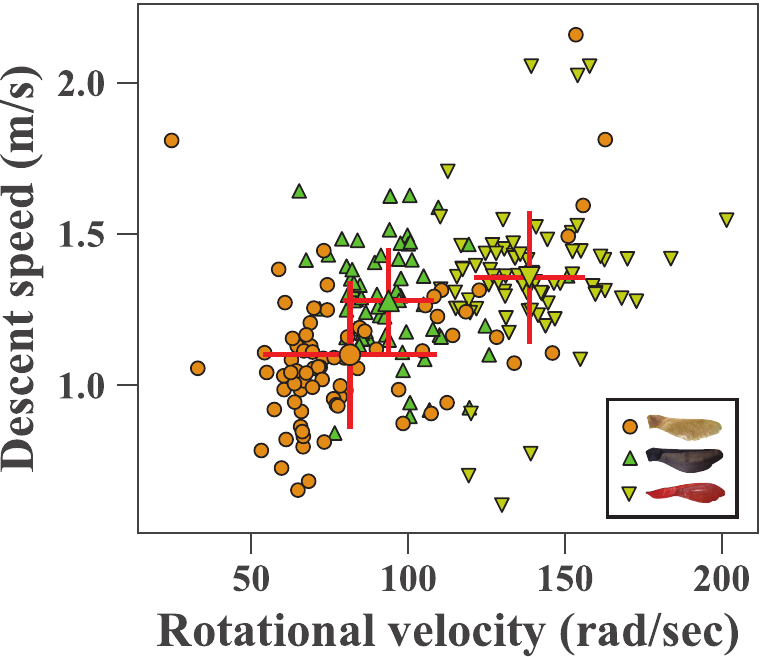} 
    &
    \includegraphics[height=0.27\linewidth]{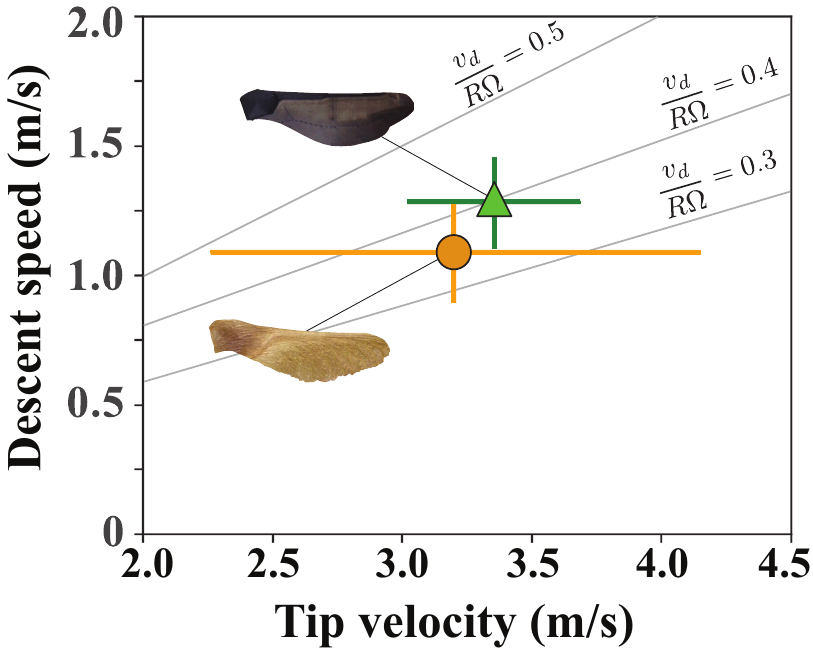} \\
    (a) & (b) & (c)
    \end{tabular}
    \caption{
    (a) Results from laboratory drop experiments in still air. The curve represents the descent path of the geometric centroid of the samara (not to be confused with the center of mass). From its rate of descent, the descent speed is calculated. From the oscillations, the rotational velocity is calculated.  A natural (left) and artificial (right) Norway maple are shown.  We have overlaid snapshots of the samaras.    
    (b) The mean descent speed ($v_d$) and rotational velocity ($\Omega$) for each trial of still air drop experiments.
    Natural Norway maple seeds are represented by orange circles, artificial  Norway maples are represented by green triangles, and artificial  silver maples are represented by yellow inverted triangles,  as shown in the legend.
    The overall mean and standard deviation of each samara type are shown by a mean point with red error bars, with shapes corresponding to each samara type. 
    (C) We show the descent speed vs.\ samara tip velocity ($R\Omega$) for the natural and artificial Norway maple. Some curves of constant descent speed to tip velocity ratio, $v_d / R \Omega$, are shown. 
    }
    \label{fig:stillAirData}
\end{figure}

Figure \ref{fig:stillAirData} shows the results of still air experiments.
In panel (b), each point represents the mean steady state descent properties, and the overall mean and standard deviation for all trials of each samara type is shown.
Taking all the samaras together, there is a positive correlation between rotational velocity descent speed, as seen previously \cite{azuma1989flight}. 
In panel (c), we show the descent speed vs.\ tip velocity, $v_t = R \Omega$, where $R$ is the mean of the distance of the samara tip from the center of mass. A few contours of the ratio of these two velocities, $v_d/R \Omega$, are shown. They are both close to $v_d/R \Omega = 0.4$, in agreement with previous work (cf.\ Fig.\ 3 of \cite{azuma1989flight}).
%Within each samara type, no clear trend exists. 
% The descent properties are additionally affected by morphological properties such as wing loading and samara shape, which will be considered below in Section \ref{ss:morphology}.

The natural Norway maple seeds exhibit both the slowest mean descent speed $v_d$ (1.10 $\pm$ 0.24 m/s) and slowest rotational velocity (81.4 $\pm$ 27.6 rad/s).
The mean properties of the  3D-printed Norway maple seeds fall within one standard deviation of their natural counterparts, with a mean descent speed of 1.28 $\pm$0 .17 m/s and a rotational velocity  of 93.7 $\pm$ 14.4 rad/s. 
Finally, the 3D-printed silver maple seeds fell more quickly (1.36 $\pm$ 0.22 m/s) and rotated more quickly (138.6 $\pm$ 17.7 rad/s) on average than either the natural or artificial Norway maple seeds. 
There was a notable variability within each samara type; the slowest descent measured was a 3D-printed silver maple, which had the fastest mean descent speed, while the fastest descent measured was a natural samara, which had the slowest mean.
The natural samaras had the most variability in both properties, likely due to increased morphological variability.
However, the variation present in the 3D-printed samaras suggests that the autorotation dynamics of even a consistent shape are highly variable when dropped from rest in still air, perhaps due to inherent sensitivity to initial conditions of the dynamics, even in a quiescent fluid medium \cite{heisinger2014coins}.

\begin{figure}
    \centering
    \includegraphics[width=\textwidth]{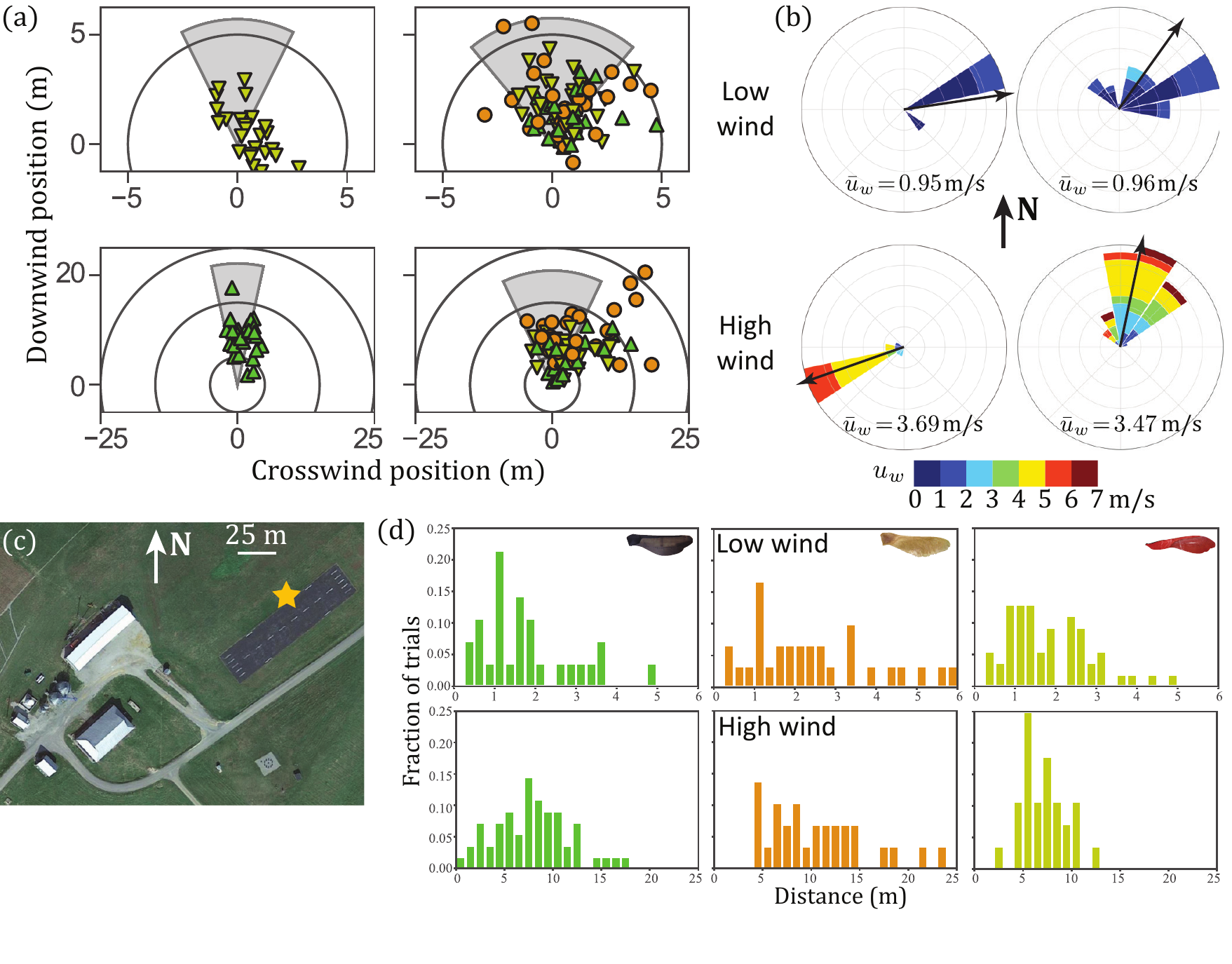}
    \caption{
    (a) Landing positions of each samara for each session of field experiments, aligned with the mean wind direction. As in Figure \ref{fig:stillAirData}, natural Norway maple seeds are shown with orange circles, artificial Norway maple seeds are shown with green triangles, and artificial silver maple seeds are shown with yellow inverted triangles.
    (b) Wind roses for each experimental session. The color of each segment represents the wind speed and the size of each segment indicates the number of occurrences. Black arrows show the mean wind direction for the session.
    (c) Aerial view of Virginia Tech field site. The release point is marked with a star and a 25 m scale bar is shown.  %Kentland experimental aerial systems laboratory. 
    Samaras were dropped from the release point at a height of 3.05 m above the runway.
    Satellite imagery: Google, Commonwealth of Virginia, Maxar Technologies.
	(d) Histograms of samara dispersal distances, for each of the wind conditions, combining the two low wind ($\approx 1$ m/s) sessions and the two high wind (3-4 m/s) sessions.
    }
    \label{fig:ODscatter}
\end{figure}

\subsection{Field experiments}

Field experiments were conducted to measure the performance (descent speed, landing location) of each samara type in natural wind conditions.
Descent speeds, measured directly from descent time, for the field drops closely aligned with the descent speeds in still air (Table \ref{tab:properties}). The natural Norway maple mean descent speeds had a difference of 7\% in the laboratory vs.\ the field setting, whereas its artificial counterpart differed by only 1\%.

The samara landing positions, and the corresponding wind rose, for each of the wind conditions, are shown in figure \ref{fig:ODscatter}.
The landing positions of figure \ref{fig:ODscatter}(a) are shown in terms of the mean wind direction, with the relative wind speed and standard deviation of the wind direction shown by the gray regions radius and angle, respectively.
The five trials which travelled the furthest in the wind were all natural samaras, and the natural samaras again showed the most variation.
Figure \ref{fig:ODscatter}(d) shows the histograms of distances travelled by each type of samara, combining the lower wind condition trials and higher wind condition trials.

\begin{figure}
    \centering
    \begin{tabular}{cc}
    \includegraphics[height=0.37\textwidth]{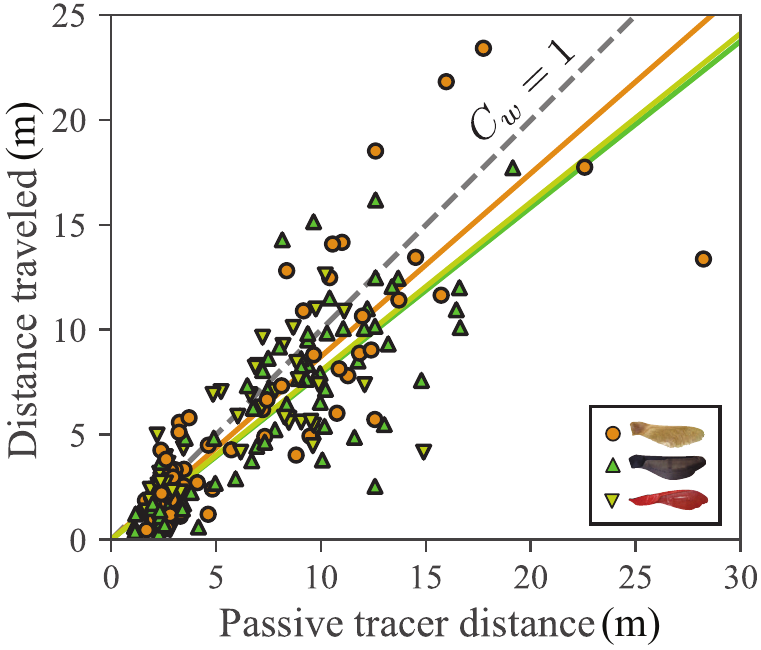} &
    \includegraphics[height=0.37\textwidth]{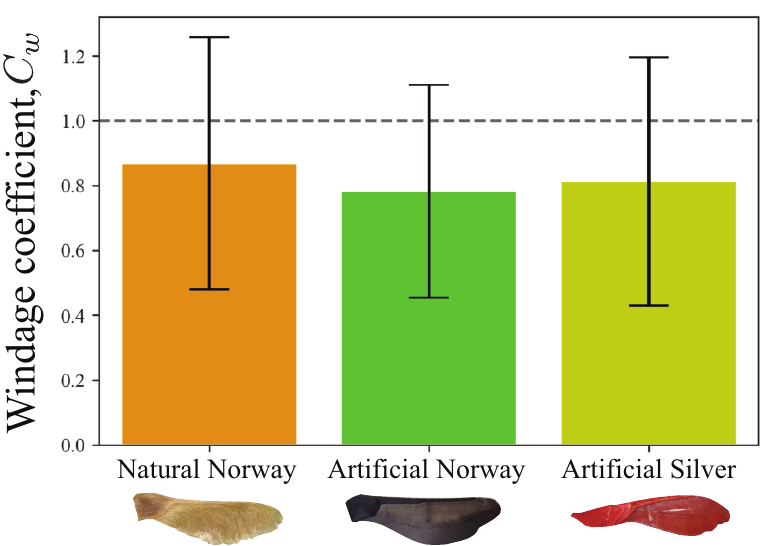}\\
    (a) & (b)
    \end{tabular}    
    \caption{
    (a) The downwind distance traveled  plotted against the effective downwind distance of a passive tracer during the same drop time. The ratio of these distances gives the windage coefficient, plotted in (b) for each of the three cases, with error bars giving standard deviation. The mean windage coefficient for each data set is shown by the straight line fit in (a).}
    \label{fig:windage}
\end{figure}

%\subsection{Windage coefficient measurements}
\subsection{Morphology as performance indicator}

\begin{figure}[h]
        \centering
        \includegraphics[width=0.8\textwidth]{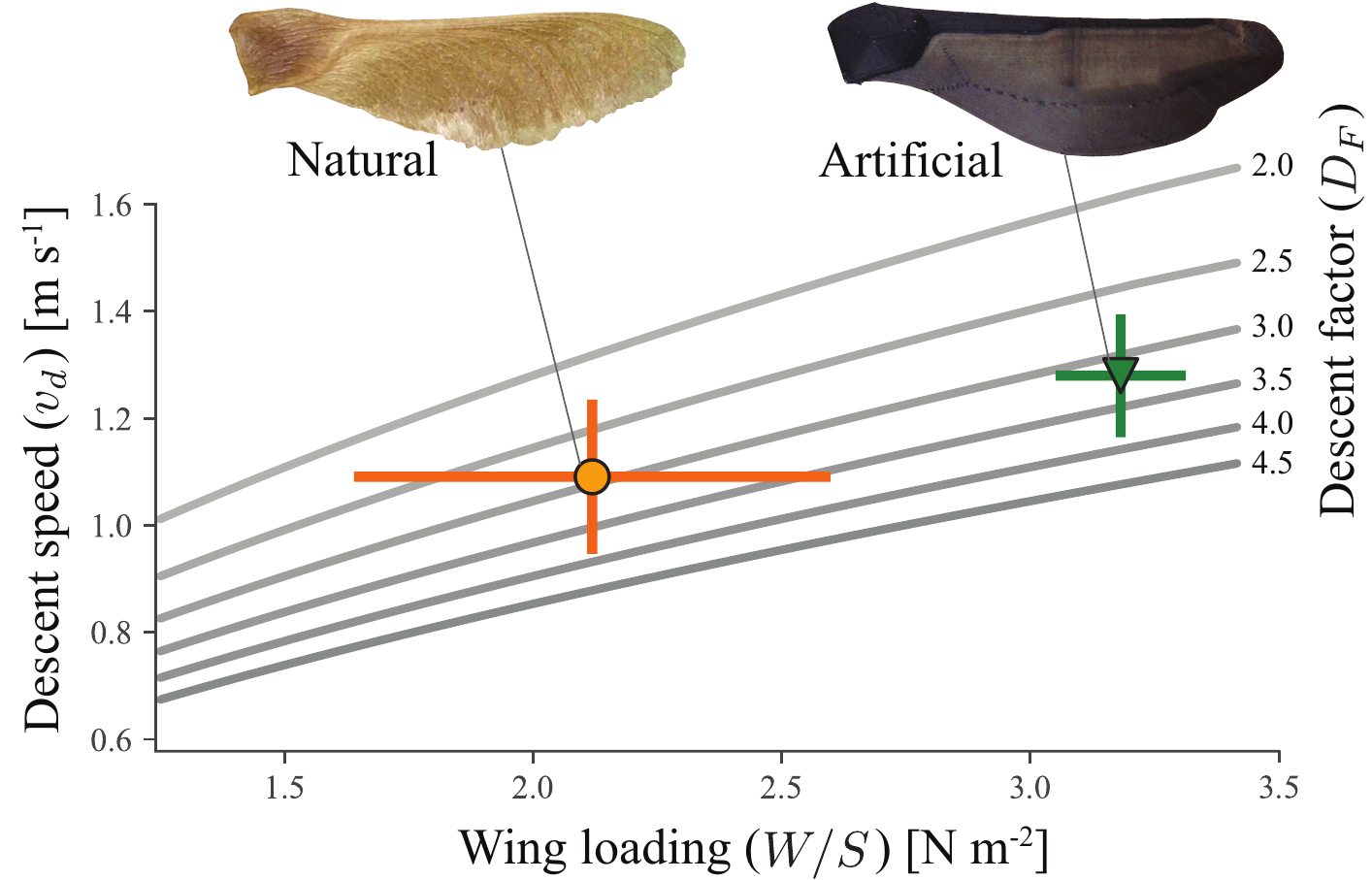}
        \caption{Descent speed ($v_d$) as a function of wing loading ($W/S$) for the natural and artificial Norway maple seeds. Mean and standard deviation are shown in both variables.
        The gray lines represent descent speed as a function of wing loading for constant descent factors ($DF$), based on (\ref{eq:lentink}), adapted from \cite{lentink2009leading}.
        The natural and artificial samaras are on the same descent factor curve, likely due to similar shapes.}
        \label{fig:lentink}
\end{figure}

Using a basic aerodynamic performance analysis, Lentink et al. \cite{lentink2009leading} derived a formula relating the descent speed to the square root of the wing loading divided by a descent factor,
\begin{equation}
v_d = \sqrt{\frac{W/S}{0.61DF}}
\label{eq:lentink}    
\end{equation}
where the descent factor $DF$ is a dimensionless number of order 1 that represents a seed's aerodynamic efficacy. In figure \ref{fig:lentink}, we plot the descent speed as a function of wing loading for both the natural and artificial Norway maple samaras. 
The gray lines represent descent speed as a function of wing loading for constant descent factors ($DF$).

\section{Discussion}

New information is needed to understand the dynamics of wind dispersal.
Here, we designed a series of biomimetic maple samaras, and tested their ability to `fly' in a series of laboratory and field experiments.
With still-air laboratory experiments, we showed that the biomimetic samaras exhibited similar descent properties (i.e., descent speed and rotation) as the natural samaras.
From field experiments, we found that the natural samara traveled the greatest distances with the wind, yet also exhibited the most variability.
Our field experiments also supported the assertion that wind speed has strong impact on dispersal.
We calculated a windage coefficient to capture the effect of wind and showed that is an effective metric for predicting dispersal distance.

The shape of the artificial samaras was developed heuristically, varying the depths of each segment until the desired behavior was observed.
However, the locations of the center of mass of the natural and artificial samaras are nearly identical.
This relationship has been more closely investigated, using balsa wood prototypes  \cite{yasuda1997autorotation}, which showed that descent speed and rotational velocity are closely related to the location of the center of mass.
The separation of the center of mass and the center of fluid forcing contributes to the rotation of a samara 
\cite{varshney2012kinematics,lee2014mechanism}.

From the observed behaviors, the natural and artificial samaras both showed approximately equivalent descent factors, $DF \approx 3$, as shown in Figure \ref{fig:lentink}.
According to \cite{lentink2009leading}, the descent factor is a measure of aerodynamic efficacy relating the wing loading, descent velocity, and fluid density of a samara.
The natural and artificial samaras having the same descent factor is likely due to their similar morphology, and suggests that differences in descent speed may be related to differences in samara weight. 

During the high wind trials, the landing distribution shows a significant bias to the right of downwind (figure \ref{fig:ODscatter}(a)).
This may be due to temporally localized turbulent gusts which were biased to the right of the mean wind direction during the trial duration (figure \ref{fig:ODscatter}(b)).
This is consistent with the high variability of the wind direction during this trial.

Our results indicate that for some samaras, the windage was estimated to be greater than 1 (see figure \ref{fig:windage}(a)).
While a windage value of greater than 1 might at first glance seem puzzling, seeds are capable of various aerial movements, including tumbling and gilding relative to the moving air \cite{rabinowitz1981dispersal,ern2012wake,nave2019global}, which could lead to average horizontal speeds which exceed that of the wind itself.
The sensitivity of these effects to local conditions, and the spatiotemporal variability of the conditions themselves (turbulence), may also contribute to the observed dispersal patterns.

{\it Biological implications.} 
A review of seed dispersal \cite{hughes1994predicting} suggests that for many plants, to escape from competition with the parent plant, there is a strong selective pressure to achieve dispersal distances of at least 1 canopy diameter. 
For a seed dropped from a canopy of height $h$ and diameter $D$, then this condition requires, via eq.\ (\ref{distance_windage}),
\begin{equation}
        C_w u_w h > v_d D . 
\end{equation}
Assuming $C_w \approx 0.8$, we have a fitness criterion of,
\begin{equation}\label{fitness_constraint}
        \frac{u_w}{v_d} > \frac{1}{C_w}\frac{D}{h}. 
\end{equation}
where we note that global average surface (10 m height) winds over land are $\approx 3$ m/s \cite{archer2005evaluation}.
For the maples considered here, we thus have $u_w/v_d \approx 3$, so the canopy diameter needs to be less than 3 times the canopy height to satisfy the hypothesized biological fitness constraint.
{\it Acer platanoides} (Norway maples) have a canopy diameter of approximately $D\approx 12$ m and a tree height of approximately $h\approx 18$ m \cite{gilman1993acer}, which gives a $C_w^{-1} D/h \approx 1.9$, satisfying the criterion, eq.\ (\ref{fitness_constraint}).

{\it Engineering implications.}
In the delivery of sensors or resources, a distributed delivery over a large area from a single release point may be a desirable feature \cite{heisinger2014coins}.
Because of the large deviation in behavior, as observed in this study, samaras naturally scatter over a range quite effectively.
Even our 3D-printed samaras, which are effectively identical morphologically, have a significant variability in their dynamic behavior both in still air in the laboratory (figure \ref{fig:stillAirData}) and in the field (figures \ref{fig:ODscatter} and \ref{fig:windage}).
We hypothesize that this diversity of behaviors is related to the stochastic nature of the transition to autorotation as well as sensitivity of samaras to local wind fluctuations (i.e., turbulence).

\section{Conclusion}
Windage factors were measured at about 80\%, which means the seeds considered are roughly moving horizontally with the wind.
Variability is higher for the natural samaras, which also show the best performance.
The windage could be measured for other wind-dispersed seeds \cite{minami2003various}, improving models that predict dispersal kernels \cite{okubo1989theoretical}, and also aiding in the design of bio-inspired delivery devices.
The ubiquity and low cost of additive manufacturing technology in the form of 3D printing make studies like this one possible and could aid in experimental tests of the windage for other seeds.
While previous studies have considered artificial samaras \cite{augspurger1987wind}, the use of 3D-printing to create replicas of seeds and study their aerodynamics can accelerate such studies.  While 3D-printing has been done previously for a double-winged seed \cite{rabault2019curving,fauli2019effect}, we note that the present study is the first, to our knowledge, to 3D-print single winged samaras. 
More detailed models and fluid experiments could be explored, especially those analyzing the side slip during autorotation \cite{norberg1973autorotation} and tumbling and gliding behaviors \cite{field1997chaotic,andersen2005unsteady}.
Another modeling direction would be the incorporation of turbulence \cite{okubo1989theoretical}, to measure if its effect on dispersal patterns, which may provide the variability observed in this study.

\section*{Acknowledgements}
This research was supported in part by a grant from the Virginia Tech Institute for Critical Technologies and Applied Sciences (ICTAS) Research Experiences for Undergraduates and grants from the National Science Foundation (NSF) under grant numbers 1821145 and 2027523.
Any opinions, findings, and conclusions or recommendations expressed in this material are those of the authors and do not necessarily reflect the views of the sponsors.

\section*{Data availability}
Files for 3D-printing are provided in the supplementary figshare repository \cite{figshare_2020}.

\bibliographystyle{shane-unsrt}
\bibliography{ross_refs3}
\end{document}